\documentclass[twocolumn,tighten]{aastex63}
\usepackage{amssymb, amsmath,framed}

\usepackage{bm}
\expandafter\ifx\csname package@font\endcsname\relax\else
 \expandafter\expandafter
 \expandafter\usepackage
 \expandafter\expandafter
 \expandafter{\csname package@font\endcsname}
\fi
\def\bq{\begin{equation}}
\def\eq{\end{equation}}
\def\bqy{\begin{eqnarray}}
\def\eqy{\end{eqnarray}}

\def\p{\partial}

\def\p{\partial}

\def\R{\mathbb{R}}

\def\calc{\mathcal{C}}

\begin{document}
\title{\large{On The Biomass Required To Produce Phosphine Detected In The Cloud Decks Of Venus}}

\correspondingauthor{Manasvi Lingam}
\email{mlingam@fit.edu}

\author{Manasvi Lingam}
\affiliation{Department of Aerospace, Physics and Space Sciences, Florida Institute of Technology, Melbourne, FL 32901, USA}
\affiliation{Institute for Theory and Computation, Harvard University, Cambridge, MA 02138, USA}

\author{Abraham Loeb}
\affiliation{Institute for Theory and Computation, Harvard University, Cambridge, MA 02138, USA}

\begin{abstract}
The detection of phosphine in the atmosphere of Venus at an abundance of $\sim 20$ ppb suggests that this gas is being generated by either indeterminate abiotic pathways or biological processes. We consider the latter possibility, and explore whether the amount of biomass required to produce the observed flux of phosphine may be reasonable. We estimate that the typical biomass densities predicted by our simple model are potentially orders of magnitude lower than the biomass density of Earth's aerial biosphere in the lower atmosphere. We briefly discuss how small spacecraft could sample the Venusian cloud decks and search for biomarkers.\\
\end{abstract}

\section{Introduction} \label{SecIntro}
It has been more than half a century since \citet{MS67} speculated about the prospects for life in the Venusian atmosphere based on its clement conditions compared to the surface; this milieu was adumbrated earlier by \citet{Sag61}.\footnote{The notion of the ``atmospheric habitable zone'' has been extended to gas giants and brown dwarfs \citep{Sha67,SS76,YPB17,LL19,LL21}.} After these pioneering papers, a number of studies have posited that the cloud layer of Venus situated at $\sim 50$ km above the surface constitute promising habitats for life because of their moderate pressures and temperatures, sufficient protection against radiation and high-energy particles, and ostensibly adequate energy and nutrient sources \citep{Grin97,SGA04,GB07,DNP15,SMI18}. On the other hand, a number of significant obstacles would have to be overcome by the hypothesized Venusian lifeforms such as combating acidity and maintaining osmoregulation \citep{Co99}.

There have been a number of notable developments in the realm of Venusian astrobiology over the past few years. Three-dimensional climate simulations suggest that Venus may have retained surface water and habitable conditions until $\sim 0.7$ Ga \citep{WDK16,WD20}, thereby providing an interval of nearly $4$ Gyr for abiogenesis to occur. The presence of strong ultraviolet absorbers in the Venusian atmosphere (whose origin remains unknown) has been long conjectured to arise from microbial activity \citep{Grin97}, and this premise was extensively investigated by \citet{LMS18} in light of recent observational data and theoretical modeling. A detailed assessment of the habitability of the Venusian atmosphere was presented in \citet{SPG20}, along with a potential life cycle for the putative Venusian microbes. \citet{IGS20} developed a broad framework akin to the famous Drake equation \citep{Dra65}, and concluded that the probability of current extant life on Venus is $\lesssim 10\%$. 

Despite these insightful theoretical analyses, no prospective gaseous biomarkers had been identified in the Venusian atmosphere. In September 2020, however, \citet{GRB20} reported the $\leq 15\,\sigma$ detection of phosphine, regarded as a candidate biosignature gas \citep{SSS20}, at a concentration of $\sim 20$ ppb in the Venusian cloud decks. The reanalysis of Pioneer-Venus Large Probe Neutral Mass Spectrometer data is also indicative of the presence of phosphine in the Venusian cloud layer at $\sim 50$-$60$ km \citep{MLW20}. It was proposed by \citet{GRB20} and \citet{BPS20} that none of the known abiotic pathways are capable of reproducing the observed abundance. Hence, if the detection of phosphine holds up to further scrutiny, there are two major possibilities to consider: either this gas is being generated by an unknown abiotic pathway or because of biological functions.

In this Letter, we examine the feasibility of the latter scenario, namely, that the detected phosphine is originating from metabolic processes. We examine the constraints on the hypothetical metabolic pathways in Sec. \ref{SecBio}, the possible biomass and average biomass density in Sec. \ref{SecPotBM}, and the prospects for \emph{in situ} detection in Sec. \ref{SecDisc}, where we also discuss some of the model limitations.

\section{Metabolic constraints}\label{SecBio}
We will describe a heuristic procedure for determining the biomass that adopts the methodology delineated in \citet[Section 3]{SBH13} - see also \citet{SKC19} and \citet{LL20} - for the so-called Type I biosignatures, which are produced when organisms extract their energy for sustenance from redox chemical gradients. The biomass in the temperate region ($\mathcal{M}_\mathrm{bio}$) can be loosely estimated as follows:
\begin{equation}\label{Mbflux}
  \mathcal{M}_\mathrm{bio} \lesssim \frac{M_\mathrm{cell}\,|\Delta G_r|\,\Phi_r \mathcal{A}}{P_\mathrm{cell}},  
\end{equation}
where $M_\mathrm{cell}$ and $P_\mathrm{cell}$ denote the characteristic mass and ``survival'' power \emph{sensu lato} of a generic putative microbe, respectively, $\Phi_r$ is the globally integrated production flux of phosphine over the cloud layer (which is inferred from observations), and $\mathcal{A} \approx 4\pi R^2$ is the total cross-sectional area, where $R \approx 0.95\,R_\oplus$ is the radius of Venus. In the above formula, $|\Delta G_r|$ embodies the Gibbs free energy available to the organism to extract from the environment in the process of producing phosphine from environmentally available chemicals. We have introduced $|\Delta G_r|$ in place of $\Delta G_r$ since the above formula is valid only for exergonic (thermodynamically favorable) reactions, i.e., when $\Delta G_r < 0$. We discuss the ramifications of this important assumption in Sec. \ref{SSecLim}. In contrast, for $\Delta G_r > 0$, we note that $\mathcal{M}_\mathrm{bio} = 0$ because energy extraction would be rendered nonviable and biomass cannot be sustained accordingly.

\begin{figure}
\includegraphics[width=7.5cm]{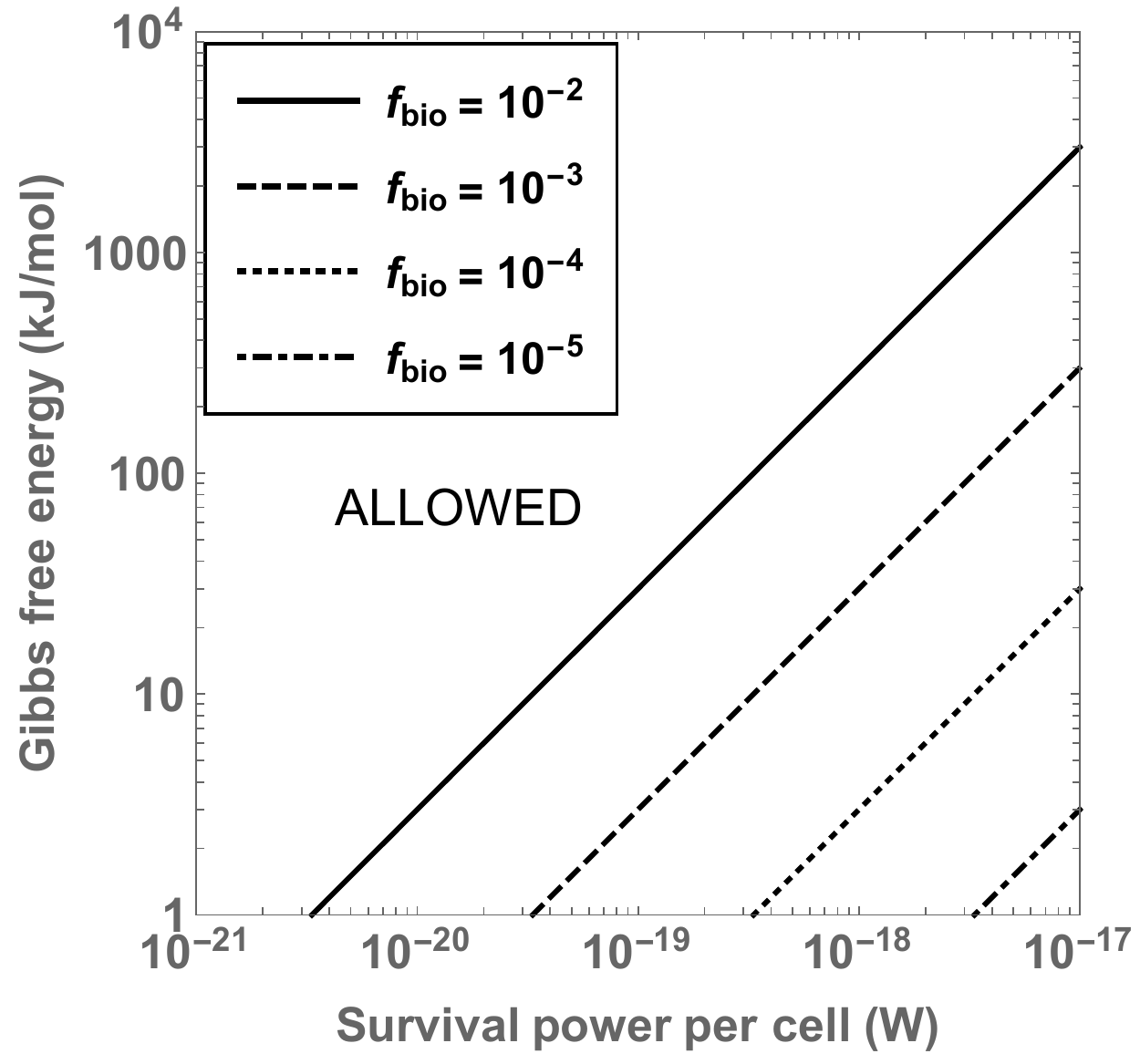} \\
\caption{The Gibbs free energy (in kJ/mol) versus the power required for survival per cell (in W) for different choices of the fraction of biogenic particles ($f_\mathrm{bio}$). The regions above the straight lines demarcate the potentially allowed parameter space.}
\label{FigMet}
\end{figure}

Of these parameters, $\mathcal{A}$ is known, and the maximal production flux was estimated to be $\Phi_r \sim \Phi_0 \equiv 10^{-13}$ mol m$^{-2}$ s$^{-1}$ by \citet{GRB20}. We are still left with a number of unspecified parameters. In order to proceed further, a couple of additional simplifying assumptions are required. It is unlikely that microbes could exist in the ``open'' for a variety of reasons, the most prominent of which is desiccation \citep{Co99,SPG20}. Hence, we will presume that the phosphine-producing microbes are embedded in droplets. Let us denote the average number density of droplets with sizes greater than $> 0.2\,\mu$m by $\eta_d$, where this limit was chosen to preserve compatibility with theoretical biophysical constraints as well as empirical data \citep{KOB99,LFW15,ABG}. We will further presume that a fraction $f_\mathrm{bio}$ of the droplets are inhabited by living microbes, and that the mean number of microbes per droplet is $\zeta_\mathrm{bio}$. In this case, the biomass is roughly given by
\begin{equation}\label{Mbaero}
 \mathcal{M}_\mathrm{bio} \sim \eta_d\, f_\mathrm{bio}\, \zeta_\mathrm{bio}\,M_\mathrm{cell}\, \mathcal{A}\, \mathcal{H},
\end{equation}
where $\mathcal{H}$ represents the height of the atmospheric habitable zone. Note that $\mathcal{V} = \mathcal{A} \mathcal{H}$ is the volume of this region. It is not unreasonable to assume that $\zeta_\mathrm{bio} \sim 1$ because the majority of particles in the temperate cloud layer are $\lesssim 1$ $\mu$m \citep{KH80,LMS18}. By equating (\ref{Mbflux}) and (\ref{Mbaero}), we end up with
\begin{equation}
    |\Delta G_r|\,\Phi_r \gtrsim \eta_d\, f_\mathrm{bio}\, \zeta_\mathrm{bio}\,P_\mathrm{cell}\,\mathcal{H}.
\end{equation}
By utilizing the simple inequality $\zeta_\mathrm{bio} \geq 1$, the above equation can be rewritten to yield
\begin{eqnarray}
&& \frac{|\Delta G_r|}{P_\mathrm{cell}\, f_\mathrm{bio}}\, \gtrsim\, 3 \times 10^{25}\,\mathrm{mol^{-1}\,s}\,\,
  \left(\frac{\mathcal{H}}{10\,\mathrm{km}}\right) \nonumber \\
&& \hspace{0.9in} \times  \left(\frac{\eta_d}{\eta_0}\right) \left(\frac{\Phi_r}{\Phi_0}\right)^{-1},
\end{eqnarray}
where we have defined $\eta_0 \equiv 3 \times 10^8$ m$^{-3}$. We have normalized $\eta_d$ as per \citet[Section 4]{LMS18} and $\mathcal{H}$ in accordance with \citet{Co99,SPG20,GRB20}. An important point worth mentioning here is that the variables on the right-hand-side (RHS) of this inequality are seemingly constrained to within an order of magnitude of their fiducial values, whereas those manifested in the left-hand-side (LHS) are unknown and therefore poorly constrained.

We have plotted this LHS in Fig. \ref{FigMet}, while holding the parameters on the RHS fixed. The regions lying above the straight lines identify the allowed parameter space. We have chosen the minimum power per cell to be $P_\mathrm{min} \equiv 10^{-21}$ W based on the synthesis of state-of-the-art theoretical models and empirical data \citep{LA15,BAA20}. The parameter space we investigate corresponds to $f_\mathrm{bio} \ll 1$ because it is anticipated that only a minuscule fraction of all the particles in the cloud layer are actually biogenic in nature \citep{SPG20}. By inspecting this figure, we see that all choices of $P_\mathrm{cell}$ yield constraints on the Gibbs free energy that might be satisfied by the conjectured metabolic pathway for sufficiently small values of $f_\mathrm{bio}$. On the other hand, if we examine the extreme limit wherein a significant fraction of all particles are biological, it is apparent that $|\Delta G_r|$ must be exceptionally high or $P_\mathrm{cell}$ must be extremely low.

\section{Potential biomass and biomass density}\label{SecPotBM}

\begin{figure*}
$$
\begin{array}{cc}
  \includegraphics[width=8.6cm]{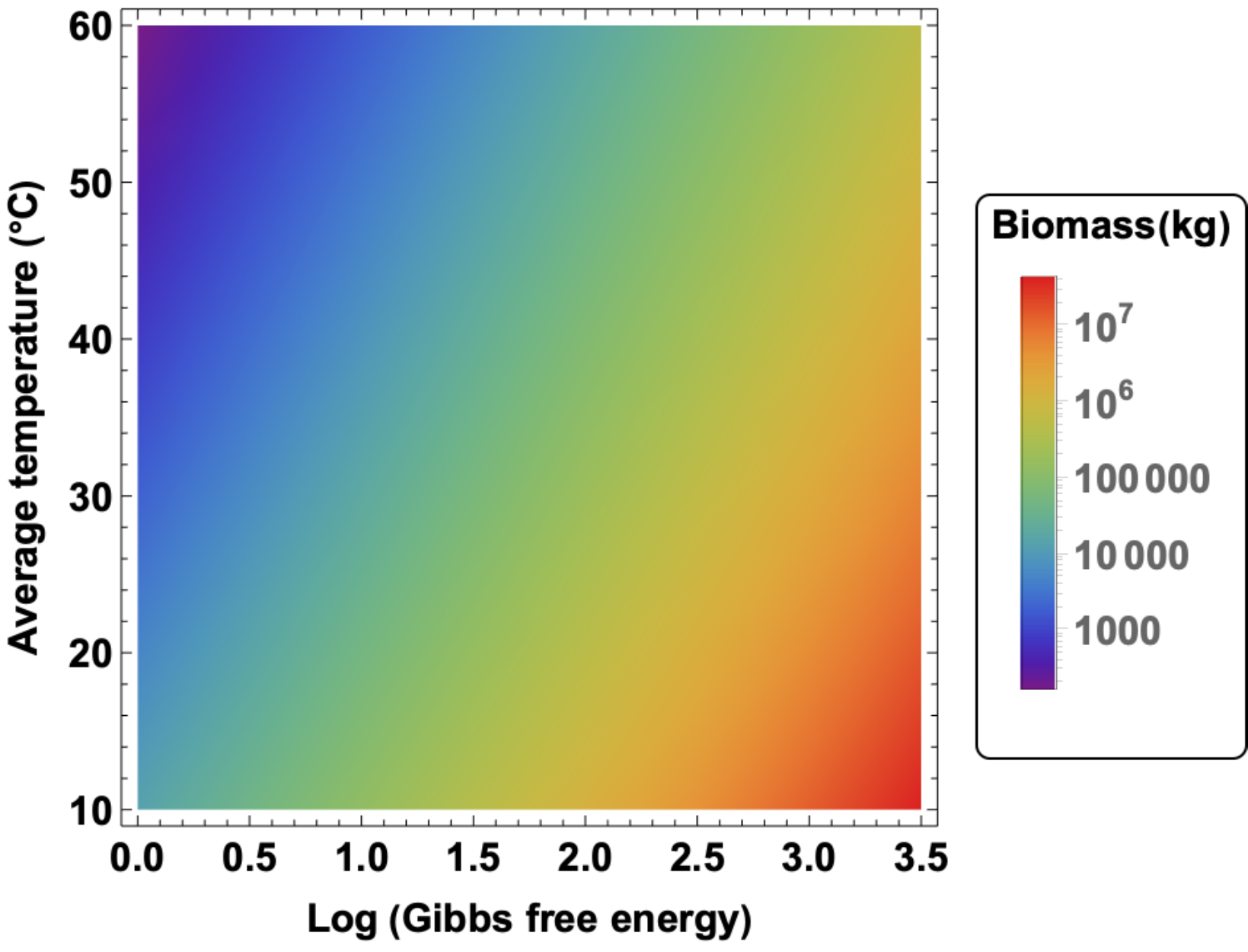} &  \includegraphics[width=9.0cm]{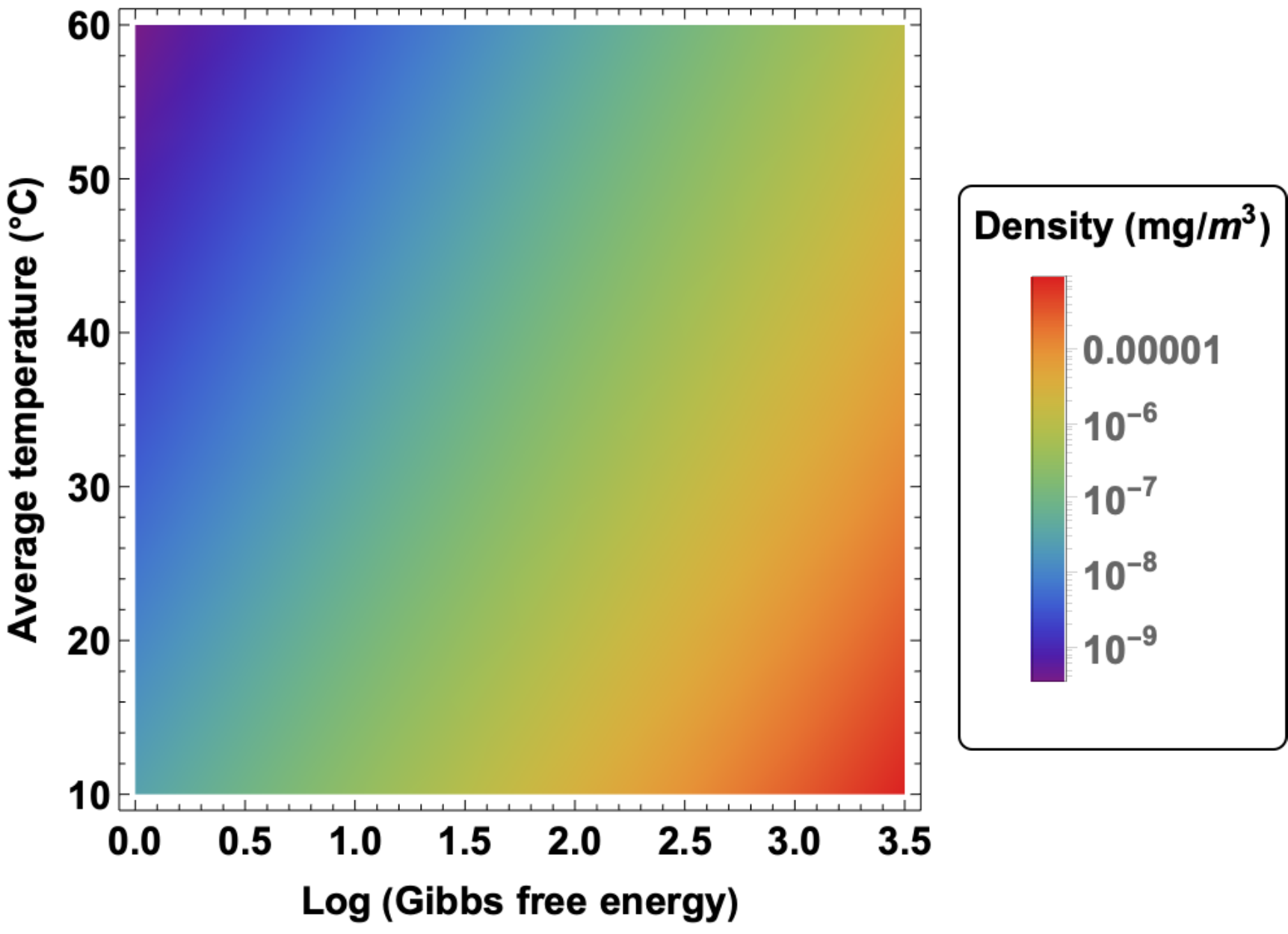}\\
\end{array}
$$
\caption{Left panel: potential biomass present in the aerial biosphere (in kg). Right panel: possible biomass density in the biosphere (in mg m$^{-3}$). In both panels, the vertical axis depicts the characteristic temperature (in $^\circ$C) of the temperate layer, while the horizontal axis quantifies the Gibbs free energy (in kJ mol$^{-1}$) for the phosphine pathway. Note the logarithmic scale on the horizontal axis, e.g., the location $2.0$ is equivalent to $100$ kJ mol$^{-1}$.}
\label{FigBioM}
\end{figure*}

We remark that prior publications have sought to estimate the mass of the Venusian aerial biosphere by developing analogs of (\ref{Mbaero}) via mass-loading calculations \citep{LMS18}. We will, instead, draw upon \citet[their Equations 11-15]{SBH13} and construct the analog of (\ref{Mbflux}) for the biomass and the average biomass density ($\rho_\mathrm{bio}$), under the assumption that the organisms are mixed throughout the region of height $\mathcal{H}$. The potential biomass $\mathcal{M}_\mathrm{bio}$ is expressible as
\begin{equation}\label{Mbiodef}
   \mathcal{M}_\mathrm{bio} \lesssim \frac{|\Delta G_r|\,\Phi_r \mathcal{A}}{\mathcal{C}}\exp\left(\frac{E_a}{R T}\right), 
\end{equation}
where $T$ could be envisioned as the mean temperature in the region where the bulk of the biomass is concentrated, while $\calc$ and $E_a$ are the survival power per unit biomass and $E_a$ is the activation energy, respectively; note that $R$ is the gas constant. Neither of the latter two parameters are known for extraterrestrial lifeforms, owing to which it is necessary to suppose that their typical values are similar to those on Earth. This assumption might be partly valid for carbon- and water-based life, but it is probably inaccurate for exotic biochemistries.

For a generic anaerobic metabolic pathway, taking our cue from the fact that the Venusian atmosphere is mostly devoid of molecular oxygen, we obtain
\begin{eqnarray}\label{Mbio}
      && \mathcal{M}_\mathrm{bio} \lesssim 2.1 \times 10^{-9}\,\mathrm{kg}\,\left(\frac{|\Delta G_r|}{1\,\mathrm{kJ/mol}}\right)\left(\frac{\Phi_r}{\Phi_0}\right) \nonumber \\
   && \hspace{0.6in}  \times \left(\frac{\mathcal{C}}{\mathcal{C}_0}\right)^{-1} \exp\left[\frac{8347}{T} \left(\frac{E_a}{E_0}\right)\right]
\end{eqnarray}
where $\mathcal{C}_0 \equiv 2.2 \times 10^{10}$ kJ kg$^{-1}$ s$^{-1}$ and $E_0 \equiv 6.94 \times 10^4$ J mol$^{-1}$ \citep{TVH93}. Next, we calculate $\rho_\mathrm{bio} = \mathcal{M}_\mathrm{bio}/\mathcal{V}$, thereby leading us to
\begin{eqnarray}\label{rhobio}
      && \rho_\mathrm{bio} \lesssim 4.5 \times 10^{-28}\,\mathrm{kg\,m^{-3}}\,\left(\frac{|\Delta G_r|}{1\,\mathrm{kJ/mol}}\right)\left(\frac{\Phi_r}{\Phi_0}\right)  \nonumber \\
   && \hspace{0.4in} \times \left(\frac{\mathcal{C}}{\mathcal{C}_0}\right)^{-1} \left(\frac{\mathcal{H}}{10\,\mathrm{km}}\right)^{-1} \exp\left[\frac{8347}{T} \left(\frac{E_a}{E_0}\right)\right].
\end{eqnarray}

We have plotted (\ref{Mbio}) and (\ref{rhobio}) in the left and right panels of Fig. \ref{FigBioM}. In both panels, the temperature range was chosen based on \citet[Section 4]{LMS18} and the Gibbs free energy range was truncated at $10^{3.5}$ kJ mol$^{-1}$, because exergonic reactions within cells on Earth rarely exceed this value \citep{Co19}. Even under the most optimal conditions specified in the figure, we end up with $\mathcal{M}_\mathrm{bio} \lesssim 4 \times 10^7$ kg and $\rho_\mathrm{bio} \lesssim 10^{-4}$ mg m$^{-3}$. Both of these estimates are not unrealistic \emph{prima facie}, in view of the harsh physicochemical conditions prevalent in the Venusian cloud decks. Furthermore, by comparing (\ref{Mbio}) with (\ref{Mbaero}) and employing our fiducial choices, we find that $f_\mathrm{bio} \ll 1$ is likely to be fulfilled.

By drawing upon the measured density of aerosols in the layer, \citet{LMS18} suggested that a maximum biomass density of $\rho_V \sim 0.1$-$100$ mg m$^{-3}$ was plausible for Venus, whereas Earth's aerial biosphere is characterized by localized regions with a density of $\rho_\mathrm{\oplus,max} \approx 44$ mg m$^{-3}$ (see \citealt{APS07}). On the other hand, the average biomass density in the lower troposphere of Earth is lower at $\rho_\oplus \sim 10^{-3}$ mg m$^{-3}$ \citep[their Table 1]{FKW16}. It is apparent that our results in Fig. \ref{FigBioM} are much smaller than these values, which indicates at the minimum that our predictions are compatible with past studies. It must be recognized, however, that our estimates for $\rho_\mathrm{bio}$ were solely for hypothetical microbes that give rise to phosphine. In the same vein, if we were to repeat the analysis for Earth using (\ref{rhobio}), it is anticipated that $\rho_\mathrm{bio}$ is negligible relative to the aforementioned densities. This ``discrepancy'' would be along expected lines, because Earth's aerial biomass is not primarily sustained by metabolic pathways producing phosphine.

We can adopt a different strategy and compute the \emph{upper bound} on the biomass ($\mathcal{M}_\mathrm{max}$) by making use of (\ref{Mbflux}) and adopting $P_\mathrm{cell} \approx P_\mathrm{min}$ for this purpose. We can compute $M_\mathrm{cell}$ by noting that most of the particles in the cloud decks have a diameter of $\sim 0.4$ $\mu$m \citep{KH80}. If roughly $50\%$ of the volume of the droplet contains the microbe, which is taken to possess a density close to that of water \citep[pp. 71-74]{MP16}, we obtain a cell mass of $\sim 1.7 \times 10^{-17}$ kg. This value is nearly identical to the average cell mass of $M_0 \equiv 1.4 \times 10^{-17}$ kg in sediments beneath the seafloor \citep{KPR12}. By substituting these numbers in (\ref{Mbflux}), the maximal biomass is estimated to be
\begin{eqnarray}\label{Mmax}
&& \mathcal{M}_\mathrm{max} \sim 6.4 \times 10^8\,\mathrm{kg}\,\left(\frac{|\Delta G_r|}{1\,\mathrm{kJ/mol}}\right) \nonumber \\
&&\hspace{0.9in} \times \left(\frac{M_\mathrm{cell}}{M_0}\right) \left(\frac{\Phi_r}{\Phi_0}\right)\left(\frac{P_\mathrm{cell}}{P_\mathrm{min}}\right).
\end{eqnarray}
As a benchmark, Earth's total biomass - which includes the aerial, surface and subsurface components - is $\mathcal{M}_\mathrm{tot,\oplus} \sim 10^{15}$ kg \citep{BPM18}, after applying the conversion from organic carbon content to biomass \citep{MFT18}. The upper bound for the biomass density ($\rho_\mathrm{max}$) derived from (\ref{Mmax}) is given by
\begin{eqnarray}\label{rhomax}
&& \rho_\mathrm{max} \sim 1.4 \times 10^{-4}\,\mathrm{mg\,m^{-3}}\,\left(\frac{|\Delta G_r|}{1\,\mathrm{kJ/mol}}\right) \left(\frac{M_\mathrm{cell}}{M_0}\right) \nonumber \\
&&\hspace{0.9in} \times \left(\frac{\Phi_r}{\Phi_0}\right)\left(\frac{P_\mathrm{cell}}{P_\mathrm{min}}\right)\left(\frac{\mathcal{H}}{10\,\mathrm{km}}\right)^{-1}.
\end{eqnarray}
If we specify $|\Delta G_r| \sim 10-100$ kJ/mol in (\ref{rhomax}), we obtain $\rho_\mathrm{max} \sim 10^{-3}-10^{-2}$ mg m$^{-3}$. Upon comparing this value against $\rho_V$ and $\rho_\mathrm{\oplus,max}$ delineated earlier, it is apparent that $\rho_\mathrm{max}$ is orders of magnitude smaller than either of the latter duo. We reiterate that this is consistent with prior publications, because we have only examined the constraints on biomass imposed by one specific (and possibly marginal) metabolic pathway. 

In closing, we note that $\mathcal{M}_\mathrm{max}$ and $\rho_\mathrm{max}$ are likely to overestimate the actual biomass and biomass density associated with this pathway by at least a few orders of magnitude. The main reason is that, even in energy-limited environments on Earth such as sediments beneath the seafloor, only a minuscule fraction ($\sim 0.02\%$) of microbes are capable of functioning at the minimal power of $P_\mathrm{cell} \approx P_\mathrm{min}$ \citep{BAA20}. In addition, at the typically warm temperatures of the cloud layer, the basal power requirements are likely to be enhanced.

\section{Discussion}\label{SecDisc}
In conclusion, we will describe some of the primary drawbacks of our modeling, and briefly explore the implications for future life-detection missions to Venus.

\subsection{Limitations of the model}\label{SSecLim}
Here, we explicate some of the salient drawbacks of the models utilized. To begin with, the framework is applicable when phosphine constitutes a metabolite in an exergonic process entailing the exploitation of chemical energy. Hence, these models are valid if and only if chemical compounds in the cloud decks of Venus exist such that the production of phosphine is favored on thermodynamic grounds, i.e., amounting to $\Delta G_r < 0$. While no such compounds are confirmed within the Venusian atmosphere at the present, there are a couple of general points that should be borne in mind.

The reduction of phosphate to phosphine is exergonic in theory at moderate temperatures provided that the environment is adequately acidic, as shown by \citet[Table 9]{BPS19} and \citet[Figure 10]{BPS20}. However, this reduction was enabled by the presence of biological substances (e.g., NADH and iron-sulfur proteins), which do not fall under the category of environmentally available chemicals. Yet, at the minimum, this datum suggests that the existence of similar reducing agents in the Venusian atmosphere ought not be dismissed \emph{a priori}. Our knowledge of the inorganic compounds in the Venusian cloud decks is largely limited \citep{TIM18}, especially when it comes to phosphorus compounds because only their gross abundance is known and not the composition of the P-containing species \citep{EBK97,Kra06}.

As both the sign and magnitude of $\Delta G_r$ are unknown in actuality, we have treated $\Delta G_r$ as a ``free'' positive parameter in our models, i.e., it constitutes the basis of our ansatz. This approach of employing an unknown parameter to encapsulate the inherent uncertainty is common to many areas of astronomy and physics ranging from the Shakura-Sunyaev viscosity in accretion discs \citep{SS73,Pri81} to dark matter and dark energy \citep{Wei08,AT10,Fe10}. We reiterate, however, that if the phosphine is generated through alternative biological processes (e.g., phototropy), the methodology would become invalid. Furthermore, the estimates for the biomass, namely (\ref{Mbflux}) or (\ref{Mbiodef}), implicitly operate under the premise of maximal efficiency. Due to the costs incurred by spillover reactions, \emph{ceteris paribus}, these formulae probably overestimate the biomass to an extent.

In the course of deriving the order-of-magnitude estimates for the relevant quantities, we ignored the spatial and temporal variability of the model parameters as well as absence of reliable knowledge (e.g., Gibbs free energy) in some instances; to bypass this issue, we drew upon Earth-based data, but this approach is not guaranteed to yield accurate results. Moreover, our treatment was exclusively centered on thermodynamic considerations, which makes the absence of \emph{kinetics} more conspicuous in view of its undoubted importance. Future studies are needed to evaluate the rate laws for the pertinent metabolic reactions, while assimilating thermodynamic effects in a self-consistent fashion \citep{JB07}.

Another crucial point deserves to be underscored here. The calculations pertaining to the biomass (or the biomass density) were exclusively predicated on putative biochemical pathways leading to phosphine production. In other words, $\mathcal{M}_\mathrm{bio}$ quantifies the amount of biomass that is linked with phosphine-generating microbes. Thus, it is \emph{not} equivalent to the total biomass, the latter of which may comprise organisms that rely on other metabolic pathways. To take the example of Earth, the biogenic flux of phosphine is $\sim 2 \times 10^{-12}$ mol m$^{-2}$ s$^{-1}$ in specialized environments \citep{SSS20,GRB20}, whereas the net global flux of O$_2$ added to the atmosphere by oxygenic photosynthesis by way of carbon burial is $\sim 6 \times 10^{-10}$ mol m$^{-2}$ s$^{-1}$ \citep{Holl02}. It is plausible, by the same token, that the total biomass on Venus might exceed our phosphine calibrated biomass $\mathcal{M}_\mathrm{bio}$ by orders of magnitude. 

Lastly, we emphasize that the preceding analysis examined a hypothetical scenario in which the phosphine is engendered due to biogenic activity. If future observations, laboratory experiments or numerical modeling demonstrate that the detection of phosphine was spurious or that it could be produced by abiotic processes, naturally our discussion will be rendered irrelevant. On account of the preceding caveats, it is necessary to interpret our results with the appropriate degree of caution. 

\subsection{Generation of phosphine via phototrophy}
We have hitherto operated under the assumption that the production of phosphine took place via exergonic processes involving chemical substances available in the Venusian atmosphere. We will now consider the opposite scenario, whereby phosphine production is endergonic in nature, i.e., $\Delta G_r > 0$. In order to drive this reaction - which is now treated as thermodynamically unfavorable (uphill) - an external energy source becomes necessary. A number of such sources are expected to exist on Venus - the most dominant among them is probably solar radiation if the example of Earth is anything to go by \citep{Deam97}, owing to which we will focus on the synthesis of phosphine by hypothetical phototrophs.

Let us suppose that the total energy flux of sunlight available in the cloud decks is $\Psi_\nu$ of which a fraction $\varepsilon$ is harnessed for metabolic purposes, thereby resulting in the production of phosphine. It is important to recognize that $\varepsilon$ encodes information about the wavelength range of the solar spectrum accessible to photoautotrophs, the conversion efficiency of light energy to chemical energy, and much more \citep{ZLO08}. A heuristic criterion for phosphine production at the detected levels is thus given by
\begin{equation}
    \varepsilon \Psi_\nu \gtrsim \Delta G_r \Phi_r,
\end{equation}
where we have used $\Delta G_r$ in place of $|\Delta G_r|$ because the reaction is now taken to be endergonic in nature. As a result, $\Delta G_r > 0$ serves as a measure of the amount of energy needed to synthesize the detected phosphine. In writing down this expression, we have implicitly presumed that phototrophy is energy-limited with $\Phi_r \propto \Psi_\nu$, but there are many other factors in reality that are capable of restricting phototrophy such as the availability of reactants, nutrients and water along with suitable temperatures \citep{LL21}. We have also supposed that the production of phosphine is continuous, and driven solely by phototrophy. After simplifying the above equation, we end up with
\begin{equation}
 \varepsilon \gtrsim 1.7 \times 10^{-13}\,\left(\frac{\Delta G_r}{1\,\mathrm{kJ/mol}}\right) \left(\frac{\Phi_r}{\Phi_0}\right)\left(\frac{\Psi_\nu}{\Psi_0}\right)^{-1},
\end{equation}
where the normalization factor of $\Psi_0 = 600$ W m$^{-2}$ is compatible with the lower bound for the total solar energy flux in the cloud decks of Venus \citep[Section 2.5]{SPG20}. In comparison, we remark that the upper bound for $\varepsilon$ is of order $0.01$ for C3 and C4 photosynthesis on Earth \citep{ZLO08}. Consider, for instance, that $\sim 1\%$ of the entire solar flux is suitable for phototrophy, that $\sim 0.01\%$ of this photon energy is used for generating phosphine via hypothetical biochemistry, and that the energy required for this process is high ($\sim 10^3$ kJ/mol). Under these restrictive conditions, the maximal production flux of phosphine is nevertheless potentially four orders of magnitude higher than $\Phi_0$. Hence, even if a tiny fraction of the radiation is harnessed by putative Venusian microbes at a very low quantum yield, the above relationship is still likely to be fulfilled.

\subsection{Sampling the Venusian atmosphere}
A number of prospective biomarkers have been identified and extensively analyzed in the context of life-detection missions \citep{SAM08,CHP19,NAD20}. Notable examples in this category include, \emph{inter alia}, enantiomeric excess in amino acids and sugars, polymers composed of nucleotides or amino acids or with repeating charges in the scaffolding, and distinct depletion of carbon and nitrogen isotopes.

Let us suppose that a fraction $\chi$ of the biological matter (organic detritus or microbes) remains viable after contact with the probe(s) equipped with instrumentation for \emph{in situ} analysis. The mass of effective biological material ($M_b$) that is collected over a time $\Delta t$ by a probe of cross-sectional area $A_p$ and travelling at relative horizontal speed $v_p$ is consequently expressible as
\begin{equation}\label{Mcoll}
 M_b \approx \chi \,\rho_\mathrm{bio}\, A_p\, v_p \,\Delta t.
\end{equation}
This relationship applies to probes that are designed to actively scoop material in the Venusian atmosphere by propelling through the cloud layer. Alternatively, balloons floating in the atmosphere could collect particles slowly drifting downward, as proposed in \citet{HLE20}. We will not examine this architecture herein. 

Given a rough estimate of $\rho_\mathrm{bio}$ and the desired $M_b$, it is feasible to tune the mission design parameters in accordance with (\ref{Mcoll}). The magnitude of $v_p$ depends on both the fuel costs incurred by braking as well as the type of instrumentation and biomarkers being sought. For example, in the case of impact ionization mass spectrometers, speeds of $v_p \sim 4$-$6$ km s$^{-1}$ are probably optimal \citep{KPH20}. One other key physicochemical constraint becomes crucial at high speeds. The bond dissociation energies of many chemical bonds and organic molecules in particular are of order $100$ kcal/mol \citep{BE03,Luo}, which translates to $\sim 4$ eV/bond. This value corresponds to the kinetic energy associated with a nucleon moving at $\sim 28$ km s$^{-1}$. At higher speeds, scooping material with a hard surface could break molecules upon impact and heat the impacting material to thousands of degrees Kelvin, thus destroying any traces of biogenic material.

When it comes to $A_p$, one can either deploy many small spacecraft akin to the Sprites on board the KickSat-2 satellite,\footnote{\url{https://www.nasa.gov/ames/kicksat}} or a few big spacecraft. Minimizing the probe size permits dispersal over a larger area, but comes at the cost of reduced flexibility vis-\`a-vis instrumentation. We will appraise the former strategy and adopt the fiducial value of $A_p \sim 0.01$ m$^{2}$, because miniature mass spectrometers are characterized by similar cross-sectional areas \citep{OC09}. By substituting these numbers in (\ref{Mcoll}), we duly end up with
\begin{eqnarray}
&& M_b \approx 3.6\,\mathrm{\mu g}\,\left(\frac{\chi}{0.01}\right)\left(\frac{\rho_\mathrm{bio}}{10^{-4}\,\mathrm{mg\,m^{-3}}}\right) \nonumber \\
&& \hspace{0.5in} \times \left(\frac{A_p}{0.01\,\mathrm{m^2}}\right)\left(\frac{v_p}{0.1\,\mathrm{km\,s^{-1}}}\right)\left(\frac{\Delta t}{1\,\mathrm{hr}}\right),
\end{eqnarray}
where we normalized $v_p$ by $0.1$ km s$^{-1}$ because dispersing this small spacecraft in the horizontal direction (from a larger one) may be feasible at this speed. The normalization for $\rho_\mathrm{bio}$ is taken to be the value described below (\ref{rhobio}); note that this choice is vastly smaller than either $\rho_V$ or $\rho_\mathrm{\oplus,max}$. For many biosignature candidates, \emph{in situ} missions require the retrieval of $\sim 1$ g of material in total \citep[Table 3]{NAD20}, but of which only a minuscule fraction (to wit, of order $\sim 1$ ng) actually has to be biological in nature. It is plausible, therefore, that small spacecraft could possess the capacity to detect biomarkers in the Venusian cloud decks even if the probes are functional over short timescales.

It should be recognized that our analysis is heuristic. Among other factors, it did not rigorously incorporate the constraints imposed by the environment (e.g., turbulence), instrument sensitivity and efficiency, and the optimal sampling of biotic material. What it does indicate, however, is that life-detection missions might yield meaningful results even if the density of microbes in the Venusian atmosphere is orders of magnitude smaller than Earth's aerial biosphere. Hence, the importance of carrying out detailed engineering studies that fulfill the desired scientific objectives is evident.

\acknowledgments
This work was supported by the Breakthrough Prize Foundation, Harvard University's Faculty of Arts and Sciences, and the Institute for Theory and Computation (ITC) at Harvard University.


\end{document}